\documentclass[aip,graphicx,preprint]{revtex4-1}

\usepackage{amsmath,amssymb,physics}
\usepackage{graphicx} 
\usepackage{dcolumn}
\usepackage{bm}

\usepackage[utf8]{inputenc}
\usepackage[T1]{fontenc}
\usepackage{mathptmx}
\usepackage{etoolbox}
\usepackage[%
    colorlinks=true,
    pdfborder={0 0 0},
    linkcolor=blue
]{hyperref}

\draft 

\makeatletter
\def\@email#1#2{%
 \endgroup
 \patchcmd{\titleblock@produce}
  {\frontmatter@RRAPformat}
  {\frontmatter@RRAPformat{\produce@RRAP{*#1\href{mailto:#2}{#2}}}\frontmatter@RRAPformat}
  {}{}
}%
\makeatother
\begin{document}

\preprint{}

\title[]{Clock Transitions Guard Against Spin Decoherence in Singlet Fission}
\author{Sina G. Lewis}
\affiliation{Department of Physics, University of Colorado Boulder,  Boulder, CO 80309, USA}
\author{Kori E. Smyser}
\author{Joel D. Eaves}
\altaffiliation{Renewable and Sustainable Energy Institute (RASEI), University of Colorado Boulder, Boulder, CO 80309, USA}
\affiliation{Department of Chemistry, University of Colorado Boulder,  Boulder, CO 80309, USA}
\email[Corresponding author: ]{joel.eaves@colorado.edu}

\date{\today}

\begin{abstract}
Short coherence times present a primary obstacle in quantum computing and sensing applications. In atomic systems, clock transitions (CTs), formed from avoided crossings in an applied Zeeman field, can substantially increase coherence times. We show how CTs can dampen intrinsic and extrinsic sources of quantum noise in molecules. Conical intersections between two periodic potentials form CTs in electron paramagnetic resonance experiments of the spin-polarized singlet fission photoproduct. We report on a pair of CTs for a two-chromophore molecule in terms of the Zeeman field strength, molecular orientation relative to the field, and molecular geometry. 
\end{abstract}


\maketitle 

\section{Introduction}
Singlet fission (SF), a photoconversion process where one photon creates two triplet excitons, has received a significant amount of attention in the literature.\cite{JosefMichl2010,JosefMichl2013} This is mostly due to its potential in next-generation solar cells to surpass the Shockley-Queisser limit.\cite{HannaNozik2006} More recently, SF has also been proposed as a method for generating qubits---or more accurately ``qudits''---for quantum applications at or near room temperature.\cite{Smyser2020,Guohua2020} 

In time-resolved electron paramagnetic resonance (trEPR) experiments, an optical pulse prepares a two-triplet excited state with singlet multiplicity $^1TT$ on fast, typically sub-nanosecond, timescales.\cite{Weiss2017,Tayebjee2017,Le2018,Lubert-Perquel2018,Chen2019} \citet{Smyser2020} showed how molecular symmetries could be harnessed to direct relaxation from $^1TT$ to specific two-triplet states with multiplicity $2S+1$ and spin angular momentum projection quantum number $M$, $^{2S+1}TT_M$. With a precise initial state defined, strong-field EPR pulses can manipulate coherences between the $^{2S+1}TT_M$ states. This scheme forms the basis for quantum logic. Used in this way, SF takes advantage of state-specific relaxation in both the internal conversion ($S_0\rightarrow\;S_n\rightarrow\; ^1TT$) and intersystem crossing ($^1TT\rightarrow \;^{2S+1}TT_M$) processes to generate highly ``spin-polarized'' states for quantum applications, solving the so-called ``state initialization'' problem.\cite{Warren1997,DiVincenzo2000}

Decoherence is the loss of phase coherence between superpositions of different quantum states. It is a major source of noise in quantum computing, information, and sensing applications. Both \emph{intrinsic} and \emph{extrinsic} fluctuations contribute to decoherence. The quantum fluctuations that produce uncertainties in identical measurements on identical systems can be viewed as a source of intrinsic noise.\cite{VanKampen} Intrinsic noise is unavoidable in molecular systems because zero-point motions deform structure. In the condensed phase, time-dependent energy gap fluctuations driven by intrinsic quantum processes---such as the polarization and vacuum fluctuations that lead to spontaneous emission and spectral diffusion---are primary sources of intrinsic noise. Intrinsic noise is often very difficult or impossible to control. By contrast, extrinsic noise can sometimes be controlled. Spin echo techniques, for example, can eliminate some extrinsic noise if it is static.\cite{Hahn1950} Examples of phenomena that contribute to extrinsic noise include the statistical distribution of energy levels at finite temperatures and the distribution of orientations of molecules in a solid.

Magnetic field fluctuations are a primary source of extrinsic noise in magnetic resonance measurements.\cite{Rosenband2008,Seo2016,Miao2020,Luis2021} Several fields in quantum measurement, from ion traps to atomic clocks and molecular magnets, take advantage of avoided crossings between various spin sublevels to dampen noise caused by fluctuating magnetic fields.\cite{Wineland2005,Rosenband2008,Friedman2020,Coronado2020,Stanton2020,Luis2021} At the avoided crossings with respect to the applied Zeeman field, transition frequencies become insensitive to field fluctuations because their first derivative with respect to the magnetic field vanishes. Transitions at these points are called ``clock transitions,'' (CTs) because of their history in atomic clocks. The dephasing ``$T_2$'' times typically show a dramatic increase around the clock transitions.\cite{Katori2003,Wineland2005,Beloy2009,Seo2016,Hill2016,Miao2020} 

Molecules, like the ones that inspire this work, have more states than atoms do. They scale better because they can keep multiple excitations coherent.\cite{Chuang2001} Unlike atoms in ion traps, however, molecules also have physical structure that results in more adiabatic variables. Unfortunately, the higher dimensional adiabatic manifolds in molecules are, in general, much more complex than they are in atoms. Molecular vibrations also introduce intrinsic noise in molecules that is manifestly absent in atoms. Finding CTs in molecules is much more challenging.

In this work, we study a model for a dimer undergoing singlet fission and find CTs in this higher dimensional adiabatic manifold. At the critical points that are analogous to the usual one-dimensional case, the first partial derivatives of the transition frequency with respect to \emph{all adiabatic variables} vanishes. The CTs stabilize transitions against intrinsic and extrinsic noise. 

In our analysis, fluctuating variables appear as adiabatic parameters in a model Hamiltonian, called the JDE model.\cite{Smyser2020} It accurately predicts trEPR spectra for the two-triplet exciton pair in strongly-coupled SF dimers. Under certain conditions, the spin-polarization from the optical preparation in the trEPR experiment is high.\cite{Smyser2020} We find a pair of CTs between the spin-polarized initial state $^5TT_0$ and the $^3TT_{-1}$ and the $^5TT_{+1}$ states. The critical points in the EPR transition energies emerge as a result of a conical intersection between the $^3TT_{-1}$ and the $^5TT_{+1}$ states on a periodic potential. We show the conditions under which the CTs can protect against decoherence from extrinsic \emph{and} intrinsic noise sources, including molecular vibrations, variations in the magnetic field, and distributions in the molecular orientations.

\section{Methods}
We employ the \emph{JDE spin exciton Hamiltonian} to describe the two-triplet exciton pair in a static Zeeman field, with the triplets from the SF process localized on chromophores $A$ and $B$.\cite{Smyser2020} Following \citet{Smyser2020}, we write the Hamiltonian in the suggestive form
\begin{equation}\label{eq:JDE_brief}
\begin{split}
    \mathcal{H} & = H_0 + H_{ZFS},\\
    & = \mu B_0 S_z - J\vec{S}_A\cdot\vec{S}_B \;+\; \vec{S_A}^\dagger\cdot\bm{D}_A\cdot\vec{S_A} \;+\; \vec{S_B}^\dagger\cdot\bm{D}_B\cdot\vec{S_B}\;,
\end{split}
\end{equation}
and choose units of angular momentum in terms of $\hbar$. The strong-field part of the Hamiltonian $H_0 = H_{Z} - J\vec{S}_A\cdot\vec{S}_B$ is a sum of the Dirac-Heisenberg isotropic exchange term $-J\vec{S}_A\cdot\vec{S}_B$ and the Zeeman interaction $H_{Z} = \mu B_0S_Z$, where $\vec{S} = \vec{S}_A + \vec{S}_B$ is the total spin and $\mu = g \mu_B$ is the magnetic dipole moment.\cite{Weil2006} The total angular momentum basis $\ket{S,M}$ diagonalizes the reference Hamiltonian $H_0$ and sets the quantization axis for all operators to the polarization direction of the Zeeman field---the $Z$ axis of the lab frame. In the literature and in this manuscript, the $\ket{S,M}$ diabatic states are also labeled by $^{2S+1}TT_M$, where $TT$ refers to the spatial nature of the spin wavefunction, the superscript denotes the state's multiplicity in terms of its total spin quantum number $S$, and $M$ is the quantum number for the projection of $\vec{S}$ along the $Z$-axis. In the field-swept EPR experiment, there is an oscillating magnetic field perpendicular to $Z$ that induces transitions between adjacent $M$ sublevels of the same $S$.

The zero-field splitting term describes the triplets interacting with themselves, $H_{ZFS} =\vec{S_A}^\dagger\cdot\bm{D}_A\cdot\vec{S_A} \;+\; \vec{S_B}^\dagger\cdot\bm{D}_B\cdot\vec{S_B}$, where $\bm{D}_A,\;\bm{D}_B$ are the spin-dipole tensors in the lab frame. In the principal frame of each chromophore---that need not coincide with one another or with the quantization axis of the magnetic field $B_0$---the spin-dipole tensor is a traceless, diagonal matrix with matrix elements $D$ and $E$. But in the lab frame, the dipole tensors are no longer diagonal and the form of $H_{ZFS}$ is more complicated (see Appendix).

\begin{figure}[t]
    \centering
    \setlength{\fboxrule}{1pt}
    \setlength{\fboxsep}{0pt}
    \fbox{\includegraphics[width=7.5cm]{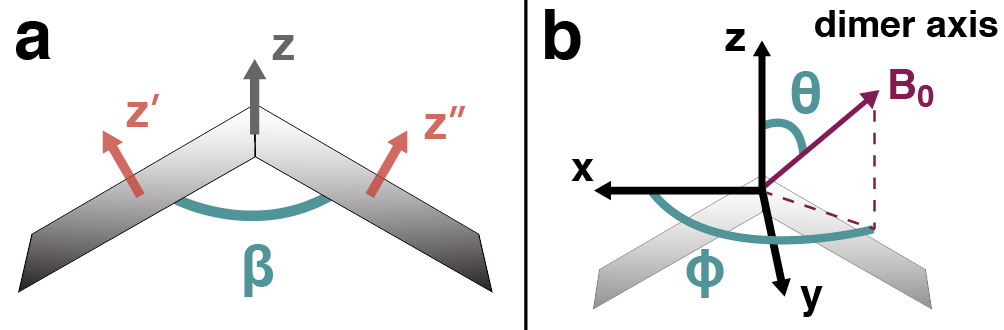}}
    \caption{Adiabatic variables include the dihedral bridging angle between two chromophores and the orientation of the magnetic field relative to the dimer. \textbf{(a)} The molecular geometry of the two chromophores (grey rectangles) joined at a bridge.\cite{Carey2018,Gilligan2019} Primed variables designate the magnetic principal axes of the individual chromophores and the unprimed ones denote the common dimer axis. \textbf{(b)} The external magnetic field $B_0$ is oriented to the dimer axis, given by the molecular geometry in \textbf{a}, by azimuthal $\phi$ and polar $\theta$ angles.}
    \label{fig:beta_phi_theta}
\end{figure}

Inspired by the pioneering synthetic work in Ref.\citenum{Carey2018} and \citenum{Gilligan2019}, we consider planar dimers connected to one another by a rigid covalent bridge that fixes their relative orientation. The angle between the chromophores is the dihedral bridging angle $\beta$ (Fig.~\ref{fig:beta_phi_theta}a), and it is the first of the adiabatic variables we consider. Molecular vibrations and structural disorder cause fluctuations in $\beta$ that contribute to intrinsic and extrinsic noise, respectively. A ``passive rotation'' relates the body frame, or dimer axis $(x,y,z)$, to the lab frame $(X,Y,Z)$, Fig.~\ref{fig:beta_phi_theta}b.\cite{Mueller2011} The details appear in the Appendix, Eq.~\ref{eq:dipoleA_rotation}-\ref{eq:dipoleB_rotation}. The orientation of the dimer frame relative to the $Z$-axis of the lab frame introduces two additional adiabatic variables, $\theta$ and $\phi$, so that $H_{ZFS} = H_{ZFS}(\beta,\theta,\phi)$. Similar to $\beta$, these variables contribute to both extrinsic and intrinsic noise. The Hamiltonian in Eq. \ref{eq:JDE_brief} also depends on the \emph{local} or effective field on a molecule \emph{in the material}. The $\bm{B}$-field in Eq. \ref{eq:JDE_brief} should really be understood as a local field. Local shielding effects make it a fluctuating variable in the ensemble which leads to decoherence.\cite{Kubo1970,Kunimasa1998} The magnetic field strength is the fourth and final adiabatic variable that we analyze in this paper, so that the total Hamiltonian depends on four adiabatic variables: $\beta$, $\theta$, $\phi$ and $B_0$. We use the shorthand $\bm{\Gamma} = (\beta,\theta,\phi,B_0)$.

The axial and rhombic EPR parameters $D$ and $E$ are fixed and we choose units of energy in terms of $J$. For Figures \ref{fig:full_energy_spectrum} and \ref{eq:directorFn}, we use $J > 0$, $D/J=0.1$, and $E=-D/4$. The sign $J>0$ orders the energies of the states from highest multiplicity to lowest, in analogy with Hund's rule for molecules, though this is not a strong rule for dimers in SF literature.\cite{Gilligan2019} With $J>0$, the exchange coupling orders the states in the same way as in Ref. \citenum{Smyser2020}. The parameter $E$ is usually small and near zero, with the opposite sign of $D$.\cite{Swenberg1973} A value $E=0$, however, might lead to accidental degeneracies in the energy spectrum. Using the maximum amplitude of $|E| = |D|/3$ avoids these spurious degeneracies. But we find it also decouples the $^1TT$ initial SF state from the rest of the $^{2S+1}TT_M$ manifold at the critical point. We instead use $E = -D/4$. These values are roughly consistent with some reported in the literature.\cite{Sternlicht1961,Yarmus1972,Weiss2017,Tayebjee2017,Lubert-Perquel2018}. Varying them quantitatively over several orders in magnitude does not change the qualitative results about the CTs that we report here.

\section{Results and Discussion}
To analyze Eq. \ref{eq:JDE_brief}, it is convenient to introduce an adiabatic and a diabatic basis. The diabatic $|S,M\rangle$ states diagonalize $H_0$. Their energies with respect to $H_0$ are only functions of $B_0/J$, $S$, and $M$. Because they are eigenfunctions of ${\vec{S}}^2$ and $S_Z$, their physical meaning is clear and they provide a convenient basis for describing state-to-state relaxation between various $^{2S+1}TT_M$ states.\cite{Smyser2020} The adiabatic states diagonalize the full Hamiltonian in the basis of the states near the crossing and their energies depend on all four of the adiabatic variables. Because they are, generally, superpositions of the diabatic states, their physical meaning is less clear.

In the space of the adibatic variables, away from degeneracies in $H_0$, the diabatic states nearly diagonalize the Hamiltonian so that there is little difference between them and the adiabatic states. But when diabatic states become nearly degenerate, there is an opportunity for a crossing. If the matrix element of $H_{ZFS}$ between the diabats near the crossing is nonzero, the adiabatic states can form an \emph{avoided crossing}. Here, the adiabatic states and energies can differ substantially from the diabats. If, however, the matrix element of $H_{ZFS}$ between the diabats is zero, the adiabats coincide with the diabats and there is a \emph{true crossing}. In the multidimensional adiabatic space of Eq. \ref{eq:JDE_brief}, there can be regions where the coupling vanishes only at a point, in which case the adiabatic curves form a \emph{conical intersection}. We find examples of all three types of crossings in this manuscript.

Avoided crossings and conical intersections are the more interesting of these three crossings in this manuscript because they can give rise to clock transitions. Dipole-allowed transitions occur at frequencies $\Omega$ and are a function of the adiabatic variables, $\Omega=\Omega(\bm{\Gamma})$. A conventional one-dimensional CT occurs at critical points $B_0^*$, where the transition frequency is insensitive to changes in the magnetic field, $\frac{d \Omega}{d B_0}|_{B_0^*} = 0$. Similarly, in our multidimensional system, CTs occur when the transition frequency is insensitive to changes in \emph{all of the adiabatic variables}, which happens at critical points $\bm{\Gamma}^*$, where  $\frac{\partial{\Omega}}{\partial \bm{\Gamma}}|_{\bm \Gamma^*} = 0$. For a clock transition to be relevant for quantum information, the initial state should be strongly spin-polarized. 

To find a relevant CT, one has to find critical points for a dipole-allowed transition that falls within the bandwidth of the EPR measurement, \emph{and} the initial state has to be strongly spin-polarized. These are rather strict requirements that might be met by a sophisticated search algorithm. We find them through a process of elimination. Avoided crossings in $B_0$ have a parabolic shape that makes them good candidates for CTs. Most of the potential crossings at $\bm{B}$-fields relevant for X-band EPR are forbidden by dipole selection rules. Other states are very weakly coupled, such as the $^1TT$ and $^5TT_{+2}$ states whose splitting is of order $(|E|/J)^2 \sim 10^{-4}$. Such weak coupling would be minimally stabilizing. This leaves two possible crossings; one between $^3TT_{-1}$ and $^5TT_{+1}$ and another between $^3TT_0$ and $^5TT_{+2}$. In this manuscript we analyze the former because the theory for the spin polarization process in the initial state is simpler. Note that this crossing would be between $^3TT_{+1}$ and $^5TT_{-1}$ for $J<0$.

\begin{figure*}[thp]
\centering
    \setlength{\fboxrule}{1pt}
    \setlength{\fboxsep}{0pt}
    \fbox{\includegraphics[width=16cm,keepaspectratio]{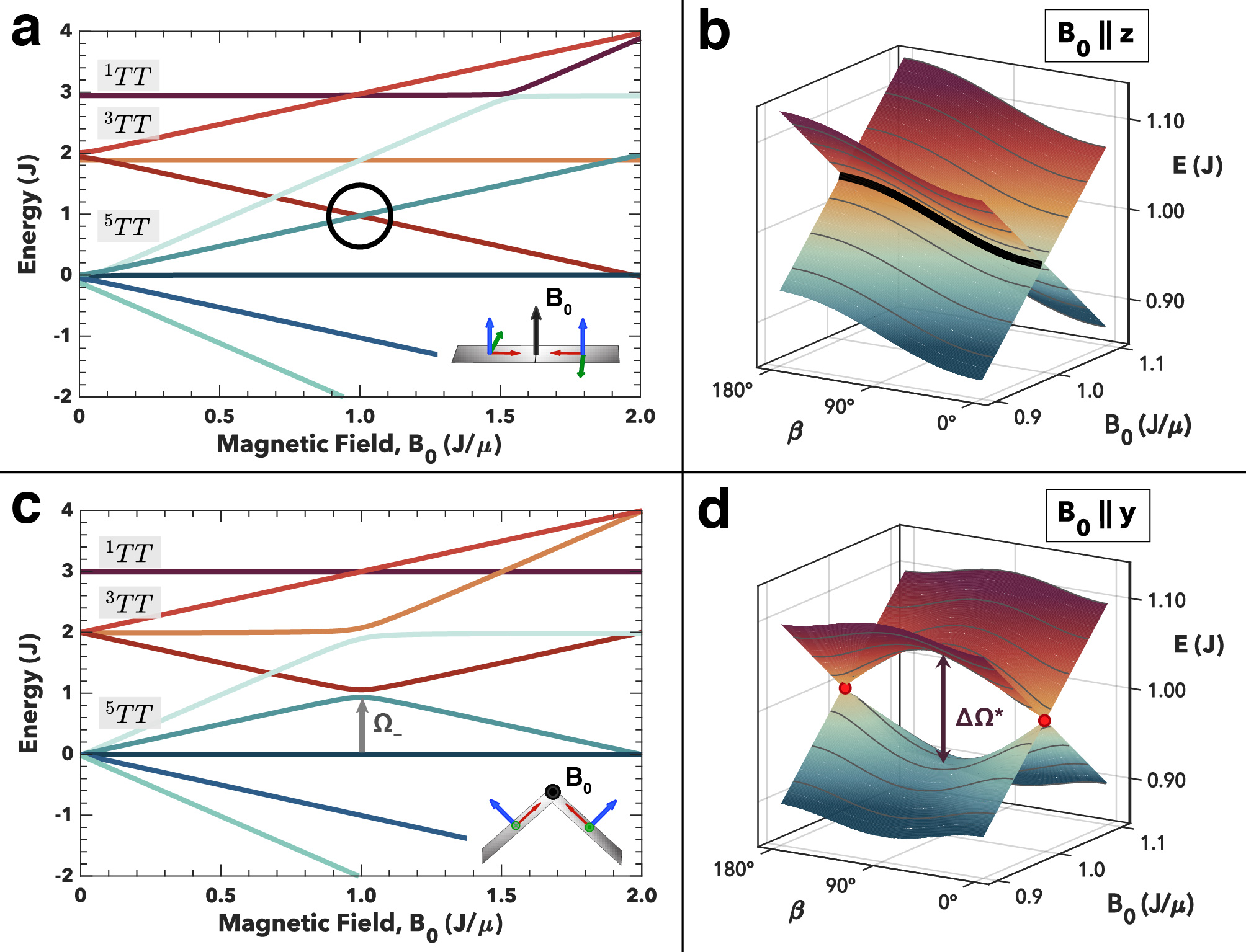}}
    \caption{The JDE Hamiltonian exhibits several potential avoided crossings, but only some are relevant to dimers in X-band EPR. The two-triplet exciton states of the singlet fission product are labeled according to their total multiplicity at strong field, $^{2S+1}TT_M$. \textbf{(a)} The field dependent energy spectrum for the all-parallel case. The black circle highlights the $S,M$ crossing we examine in (a-d). \textbf{(b)} Upper and lower adiabatic states in the $(\beta,B_0)$ plane with $B_0 \parallel z$. There is a line of true crossings (black line) on resonance, $\delta = 0$ for all $\beta$. \textbf{(c)} The field dependent energy spectrum for $B_0\parallel y$ and $\beta=90^\circ$. The grey arrow in (c) indicates the transition stabilized by the saddle-point in (d). \textbf{(d)} Upper and lower adiabatic states in the $(\beta,B_0)$ plane with $B_0 \parallel z$ and $\beta = 90^\circ$. The coupling only disappears at a single point, where the chromophores are parallel (red points). This forms a conical intersection in the $(\beta, B_0)$ plane, in contrast to the true crossings diagrammed in (b). The saddle-points in (d), indicated by the double arrow that shows the maximal splitting, $\Delta \Omega^* = \Delta \Omega({\bm{\Gamma}}^*)$, away from the conical intersections can stabilize the system against fluctuations in $\beta$ and $B_0$. The transitions corresponding to $\Omega_+$ and $\Omega_-$ are the \textit{clock transitions}.}
    \label{fig:full_energy_spectrum}
\end{figure*}

In the trEPR experiment, a laser pulse initiates the SF process which populates the singlet $^1TT$ level.\cite{Gilligan2019} Provided that $J$ is large, fluctuations in $J$ drive non-adiabatic transitions to other levels in the $^{2S+1}TT_M$ manifold.\cite{Smyser2020} But the triangle inequality, resulting from the Wigner-Eckart theorem (Appendix Eq.~\ref{eq:triangle_inequality}) forbids relaxation from the singlet $^1TT$ state to the triplet $^3TT_M$ levels. This has two consequences. First, all crossings between $^1TT$ and $^3TT_M$ are true crossings (Fig.~\ref{fig:full_energy_spectrum}a,c). Second, the relaxation from $^{1}TT$ populates only the quintet $^5TT_M$ levels with a rate proportional to $|\langle ^1TT|H_{ZFS}|^{5}TT_M\rangle|^2$. These matrix elements determine the selection rules and the spin-polarization in the initial state. If they are large for particular values of $M$ and small for others, the relaxation is state-specific and the quintet is strongly spin-polarized.\cite{Smyser2020} Because $H_{ZFS}$ depends on the adiabatic variables, the strength of the spin-polarization does too.

\citet{Smyser2020} showed that the ``all-parallel'' case of $\beta=180^\circ$ with the magnetic field aligned along the dimer $z-$axis ($B_0 \parallel z$) generates strong spin-polarization in $^5TT_{0}$. From this state, there are only a few crossings that are relevant to dimers in X-band EPR. The nearby crossing circled in Figure~\ref{fig:full_energy_spectrum}a is the only one allowed by selection rules whose transition is dipole-allowed. The transition between this state and the initial spin-polarized $^5TT_0$ state is a candidate CT. 

The energy spectrum of the adiabats for the all-parallel case appears in Fig.~\ref{fig:full_energy_spectrum}a. Exchange symmetry demands that the couplings between the triplet and quintet levels vanish for all values of $\beta$. It follows that all crossings are true crossings in the $(\beta,B_0)$ plane (Fig.~\ref{fig:full_energy_spectrum}b).\cite{Smyser2020} Similarly, the crossing remains true for all values of $\phi$ and $\theta$ when $\beta=180^\circ$. The true crossing between the quintet and triplet states precludes CTs under these conditions.

When the molecule is bent ($\beta \neq 180^\circ$) \emph{and} the magnetic field is not directed along the $z-$axis of the chromophore pair, the intersection between $^3TT_{-1}$ and $^5TT_{+1}$ is no longer a true crossing. It now becomes a candidate for a clock transition. Near this crossing, we define the upper or lower energy branch of the adiabatic states as $E_-$ and $E_+$, respectively. Intensity borrowing from $^5TT_{+1}$ allows transitions from $^5TT_0$ to both the upper and lower branch. Shifting the reference energy so that the $^5TT_0$ energy is zero, the transition frequencies are $\Omega_{\pm} = E_\pm/(D-E)$. $\Omega_-$ is shown as a grey arrow in Fig.~\ref{fig:full_energy_spectrum}c. A useful rearrangement of $\Omega_{\pm}$ gives $\Omega_{\pm}= \overline\Omega\;\pm\;\Delta\Omega$, where $\overline\Omega$ (Eq. \ref{eq:Omega_tilde}) is a carrier frequency.

Figure~\ref{fig:full_energy_spectrum}d shows the results for the upper and lower adiabtic branches in the $(\beta,B_0)$ plane for $B_0 \parallel y$. As before, the coupling vanishes when chromophores are parallel, at $\beta = 180^\circ$, but is nonzero for all other values. The upper and lower branches form a conical intersection. The periodic nature of the bridging angle imposes a boundary condition on the surfaces as they move away from the conical intersection. The lower and upper branch must have a critical point away from the conical intersection because the conical intersection pins both surfaces at the parallel $\beta = 0$ and the anti-parallel $\beta = 180^\circ$ geometries (Fig. \ref{fig:full_energy_spectrum}d). 

At the conical intersection, the transition frequency would change violently in the space of the adiabatic variables. But, away from the conical intersection, there is a pair of saddle points in the $(\beta,B_0)$ plane that can stabilize a pair of transitions. As we will show, these saddle points do indeed reflect a critical point in the full dimensional adiabatic space and correspond to clock transitions.

The transition frequencies are split by the difference $\Delta \Omega = \frac{\Omega_+ - \Omega_-}{2}$, which depends on all four adiabatic variables. It can be written in the form
\begin{align}
    \Delta\Omega &= \sqrt{\delta^2 + \sin^2(\beta)f(\phi,\theta)}\;,\label{eq:Omega}
\end{align}
where
\begin{align}
    f(\phi,\theta) &= \frac{1}{4}\sin^2(\theta)\left(1-\cos^2(\phi)\sin^2(\theta)\right).\label{eq:directorFn}
\end{align}
Candidate clock transitions occur for nonzero values of $\Delta\Omega$. More specifically, $\Delta \Omega$ is the splitting between the pair of clock transitions, which must be larger than the linewidths $\sim 1/T_2$ for the clock transitions to be resolvable. In Eq. \ref{eq:Omega}, $\delta \equiv (\mu B_0 - J)/(D-E)$ is a more convenient variable than $B_0$ (Fig~\ref{fig:full_energy_spectrum}). But $\delta$ neatly expresses the detuning of $B_0$ from the avoided crossing, which occurs at resonance, $\delta=0$ (black circle Fig.~\ref{fig:full_energy_spectrum}a). The function $f(\phi,\theta)$ is what we call \emph{the orientor} because it completely describes the orientational dependence of the CT. In avoided crossings between two-level systems, finding the adiabatic states that diagonalize the Hamiltonian is equivalent to choosing a mixing angle that depends on two scalars. From this perspective Eq.~\ref{eq:Omega} describes the transition frequency for a two-level system whose mixing angle depends on several variables $\delta$, $\beta$, $\phi$, and $\theta$.

We find critical points of Eq.~\ref{eq:Omega} analytically---it turns out that critical points in $\Delta\Omega=\Delta\Omega^*$ (double arrow Fig.~\ref{fig:full_energy_spectrum}d) are also critical points in $\Omega_\pm$. At resonance, $\Delta\Omega$ is a completely separable function of the bridging angle $\beta$ and the magnetic field orientations $\phi$ and $\theta$. This greatly simplifies the analysis. 

At resonance, the dependence on the bridging angle is a multiplicative factor of $\sin(\beta)$. This implies that a dimer with a bend of $90^\circ$ would minimize extrinsic fluctuations due to orientational disorder and intrinsic fluctuations from molecular vibrations that modulate the orientations. 

\begin{figure}[ht]
    \centering
    \setlength{\fboxrule}{1pt}
    \setlength{\fboxsep}{0pt}
    \fbox{\includegraphics[width=7.5cm]{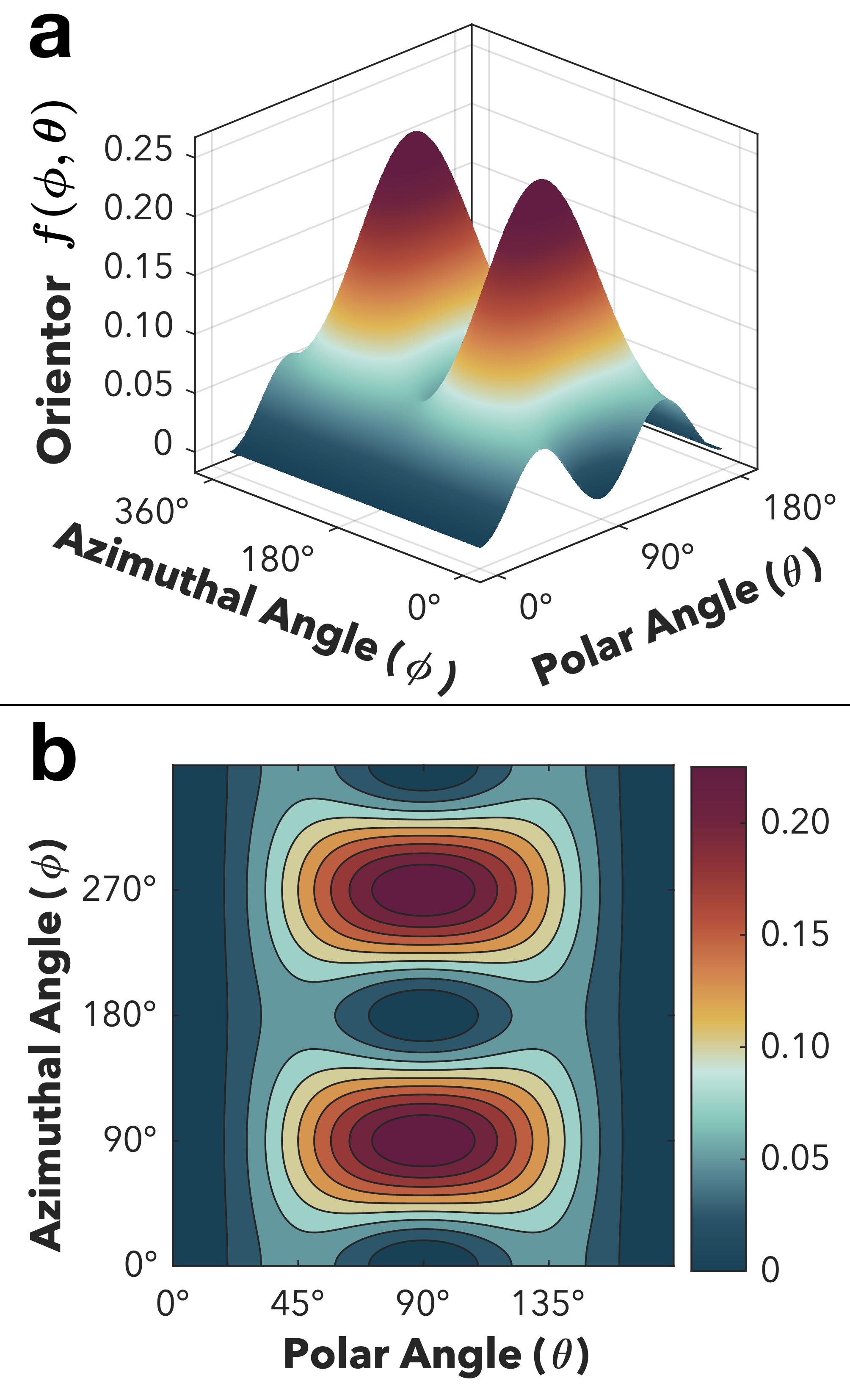}}
    \caption{The orientor function, $f(\phi,\theta)$ completely describes the orientational dependence of the splitting between upper and lower adiabatic states, given by $\Delta\Omega$ on resonance ($\delta = 0$) when the chromophores are bent ($\beta = 90^\circ$). These two adiabatic variables are set to the saddle point of Fig.~\ref{fig:full_energy_spectrum}d. \textbf{(a)} Surface plot and \textbf{(b)} contour plot of the orientor. The splitting $\Delta\Omega=\sqrt{f(\phi,\theta)}$ depends only on the orientation of the field relative to the molecular axes $\phi,\;\theta$ (Fig~\ref{fig:beta_phi_theta}b). The maxima of the orientor $f(\phi,\theta)$ are critical points that stabilize the transition frequencies against spatial fluctuations caused by static disorder or nuclear vibrations in the angles $\phi,\;\theta$ of the magnetic field relative to the dimer axis.}
    \label{fig:theta_phi_plot}
\end{figure}

The orientor is a more complex function (Fig. \ref{fig:theta_phi_plot}). It has maxima at the critical points $\Gamma^*$ when $B_0 \parallel y$, so that $B_0$ is aligned with the individual chromophore $y'$ and $y''$ axes, which are parallel to one another for any value of $\beta$. The orientor also has a pair of saddle points at $(\phi,\;\theta) = (0^\circ,45^\circ)$ and $(\phi,\;\theta) = (180^\circ,45^\circ)$ that we disregard. The $^5TT_0$ state is weakly spin-polarized and $\Delta \Omega \sim \sqrt{f}$ is small at those saddle points compared to the maxima in $f$. The alignment of $B_0\parallel y$ minimizes extrinsic sources of noise resulting from the distribution of orientation in the material as well as intrinsic noise from nuclear vibrations that modulate the orientation. Note that when $B_0 \parallel z$, the orientor is zero and the transition is not stabilized. 

\section{Conclusion}
Clock transitions have a history in atomic physics literature. For quantum information applications, molecules have several advantages over atoms, but with a cost of increased complexity. We analyze the JDE model, Eq. \ref{eq:JDE_brief},\cite{Smyser2020} to predict a pair of clock transitions for two-triplet excitons on rigid, covalently bonded, planar dimers. Choosing the diabatic basis with respect to the lab axis expresses the entire Hamiltonian in the stationary states of the Zeeman Hamiltonian. Away from the crossings, the zero-field Hamiltonian, while complicated in this basis, is perturbatively small. This choice of basis also provides a consistent framework for determining spin polarization in the initial states when $J$ is large.\cite{Smyser2020} 

The adiabatic eigenstates of the full Hamiltonian, as a function of the four adiabatic variables that we analyze, contain three types of crossings: avoided crossings, true crossings, and conical intersections. Under certain conditions, avoided crossings and conical intersections give rise to clock transitions. Dipole selection rules in concert with mixing rules derived from the Wigner-Eckhart theorem (see Appendix), imply that only a few crossings are relevant candidate CTs for X-band EPR. 

For a relevant CT, the initial state must also be spin-polarized. We find a pair of CTs from the $^5TT_0$ state and the upper and lower adiabatic branches resulting from the avoided crossing between $^3TT_{-1}$ and $^5TT_{+1}$. Transitions to both branches are stabilized at the critical point $\bm{\Gamma}^* = (\beta=90^\circ,\theta=90^\circ,\phi=90^\circ,B_0=J/\mu)$---a high symmetry configuration where the magnetic field orientation $B_0\parallel y$, aligns $B_0$ with the chromophore $y'$ and $y''$ axes for all values of $\beta$ (Fig.~\ref{fig:full_energy_spectrum}c inset).

Compared to atomic systems, molecules offer many advantages for quantum applications. Molecules, however, also have more intrinsic noise that is often difficult to control. A molecule with three dimensional structure has a non-trivial relationship between the body frame and the lab frame. The parameters that characterize this relationship introduce additional adiabatic variables that we analyze here. Internal molecular structure---in this work, the dihedral bridging angle $\beta$---introduces new adiabatic variables too. We find a pair of clock transitions for SF chromophores that, remarkably, satisfy all criteria required of a clock transition simultaneously. Clock transitions offer a rare opportunity to control intrinsic noise. We have predicted a pair of clock transitions that can extend decoherence times for two-triplet excitons on rigid, covalently bonded, planar dimers. While the molecules that we suggest have not, to our knowledge, yet been synthesized, we hope that our work inspires their creation.

\section*{Author Contributions}
JDE designed the research. JDE, SGL, and KES performed the research and wrote the manuscript.

\begin{acknowledgments}
We thank Obadiah Reid and Brandon Rugg for introducing us to the topic of clock transitions. Funding was provided by the United States Department of Energy, Office of Basic Energy Sciences (ERW7404).
\end{acknowledgments}

\section*{Competing Interests}
The authors declare no competing interests.

\section*{Data Availability Statement}
The data that support the findings of this study are available from the corresponding author upon reasonable request.

\appendix

\section{Appendix A: Selection Rules of the JDE Model}
\renewcommand{\theequation}{A.\arabic{equation}}
\setcounter{equation}{0}
The reference Hamiltonian is diagonal in the diabatic basis discussed in the text
\begin{equation}
    \bra{S',\;M'}H_0\ket{S,\;M} = \left(\mu B_0 M  - \frac{J}{2}(S(S+1) - 4)\right)\delta_{S'=S}\delta_{M'=M}.
\end{equation}
Eq.~\ref{eq:JDE_brief} is written in the lab frame. Solving for the matrix elements of the zero-field splitting Hamiltonian requires relating the spin-dipole tensor in the principal frame, where the parameters $D$ and $E$ are known, to its representation in the lab frame. In the principal frame of chromophore $i$ the spin-dipole tensor $\bm{D_i}''$ is symmetric and traceless
\begin{eqnarray}
    \bm{D_A}'' = \bm{D_B}'' = \begin{pmatrix}
    -D/3 + E & 0 & 0\\
    0 & -D/3-E & 0\\
    0 & 0 & 2D/3
    \end{pmatrix}.
\end{eqnarray}
We relate the principal frame of the two chromophores to the lab frame by a series of rotations. First, the molecules are symmetrically rotated via an active rotation $R_A(\pm\Theta_\beta)$ to obtain the desired bridging angle $\beta$ of the dimer as shown in Fig. \ref{fig:beta_phi_theta}a. Second, a passive rotation $R_P(\Theta_B)$ is applied to both chromophores to describe the orientation of the applied magnetic field with respect to the shared dimer axis.
\begin{eqnarray}
    \bm{D_A} = \bm{R}_P(\Theta_B) \bm{R}_A(-\Theta_\beta)\bm{D_A}''\bm{R}_A^{-1}(-\Theta_\beta)\bm{R}_P^{-1}(\Theta_B),\label{eq:rotationsDA}\label{eq:dipoleA_rotation}\\
    \bm{D_B} = \bm{R}_P(\Theta_B) \bm{R}_A(\Theta_\beta)\bm{D_B}''\bm{R}_A^{-1}(\Theta_\beta)\bm{R}_P^{-1}(\Theta_B).\label{eq:rotationsDB}\label{eq:dipoleB_rotation}
\end{eqnarray}
Following the method outlined in Mueller\cite{Mueller2011}, we write the matrix elements of $H_{ZFS}$ as
\begin{widetext}
\begin{eqnarray}
    \bra{S'\;M'}H_{ZFS}\ket{S\;M} = \sum_{i,j,k=-2}^{2} (-1)^{k-M'}\sqrt{5(2S+1)(2S'+1)}
    D_i^{(2)}\left(\mathcal{D}_{kj}^{(2)}(\Theta_B)\right)^\dagger\nonumber\\
    \times\begin{pmatrix}
    2 & S & S'\\
    -k & M & -M'
    \end{pmatrix}\begin{Bmatrix}
    1 & 1 & 2\\
    S' & S & 1
    \end{Bmatrix}\left(\mathcal{D}_{ji}^{(2)}(\Theta_\beta) + (-1)^{S'+S}\mathcal{D}_{ji}^{(2)}(-\Theta_\beta)\right)\;,
\end{eqnarray}
\end{widetext}
where $D_i^{(2)}$ is the $i$th component of the 2nd rank spherical tensor representation of the ZFS tensor $\bm{D}$ in the principal frame (e.g. $D_0^{(2)} = \sqrt{2/3}D,\;D_{\pm1}^{(2)} = 0,\;D_{\pm2}^{(2)} = E$) and $\mathcal{D}_{i,j}^{(2)}(\Theta)$ are the Wigner rotation matrix elements for a rotation defined by the set of angles $\Theta$. We use the typical physics convention to define the Wigner rotation matrix elements. The Wigner-3j symbol is best defined in relation to Clebsch-Gordon coefficients for adding angular momentum $j_1$ and $j_2$ to get $j_3$,
\begin{align}
    \begin{pmatrix}
    j_1 & j_2 & j_3\\
    m_1 & m_2 & m_3
    \end{pmatrix} &= \frac{(-1)^{j_2-j_1+m_3}}{\sqrt{2 j_3 + 1}}\braket{j_1, m_1; j_2, m_2}{j_1,j_2;j_3,-m_3}.
\end{align}
The Wigner-6j symbol is a sum over Wigner-3j symbols,
\begin{align}
    \begin{Bmatrix}
    j_1 & j_2 & j_3\\
    j_4 & j_5 & j_6
    \end{Bmatrix} &= \sum_{m_i}(-1)^{\sum_k(j_k - m_k)}
        \begin{pmatrix}
        j_1 & j_2 & j_3\\
        -m_1 & -m_2 & -m_3
        \end{pmatrix}\nonumber\\[0.15cm]
        &\hspace{-1cm}\times
        \begin{pmatrix}
        j_1 & j_5 & j_6\\
        m_1 & -m_5 & -m_6
        \end{pmatrix}
        \begin{pmatrix}
        j_4 & j_2 & j_6\\
        m_4 & m_2 & -m_6
        \end{pmatrix}
        \begin{pmatrix}
        j_4 & j_5 & j_3\\
        -m_4 & m_5 & m_3
        \end{pmatrix}.
\end{align}
The orthogonality properties of the Wigner-3j and Wigner-6j symbols provide the selection rules for transitions between $\ket{S',\;M'}$ and $\ket{S,\;M}$
\begin{align}
   S'+S &\ge 2, \label{eq:spin_addition_rule}\\
   |S-2| \le &\;S' \le S+2, \label{eq:triangle_inequality}\\
   M-M' &= k, \label{eq:M_addition_rule}
\end{align}
where $k$ is an integer between $-2$ and $2$. Eq.~\ref{eq:spin_addition_rule} follows from the Wigner-6j triangle inequality $|j_4-j_5|\le j_3\le j_4+j_5$. Eq.~\ref{eq:triangle_inequality}, which follows from the Wigner-3j triangle inequality, can be shown to give the same restriction.
We note that for parallel chromophores, where $\mathcal{D}_{ji}^{(2)}(\theta_\beta)=\mathcal{D}_{ji}^{(2)}(-\theta_\beta)$, we require the additional condition $S' + S$ is even. For parallel chromophores, the coupling matrix elements between $^3TT$ and $^1TT,\;^5TT$ are strictly zero.

\section{Appendix B: Transition Frequency between Quintet $M=0$ and Upper/Lower Adiabatic Branch}
With the reference energy set so that the $^5TT_0$ energy is zero, the transition frequency to the adiabatic states considered in the main text $\Omega_{\pm} = E_\pm/(D-E)$ is
\begin{align}
    \Omega_{\pm} &= \overline\Omega\;\pm\;\Delta\Omega,
\end{align}
where
\begin{align}
    \overline\Omega &=1+\frac{(D+3E)}{12}(1-3\sin^2\theta\sin^2\phi)\nonumber\\
    &\hspace{1.5cm} +\frac{(D-E)}{4}\cos\beta(\sin^2\theta(1+\cos^2\phi)-1)\label{eq:Omega_tilde},\\
    \Delta\Omega &= \sqrt{\delta^2 + \sin^2(\beta)f(\phi,\theta)},\\
    f(\phi,\theta) &= \frac{1}{4}\sin^2(\theta)\left(1-\cos^2(\phi)\sin^2(\theta)\right),
\end{align}
and $\delta = (\mu B_0 - J)/(D-E)$.

\bibliography{biblio.bib}

\end{document}